\documentstyle[12pt,a4]{article}
\begin{document}
\centerline{\Large \bf On Quantum Deformations of $d=4$ Conformal Algebra
\footnote{Partially supported by KBN grant 2P03B130.12}
\footnote{To appear in the proceedings of the 7$^{th}$ Colloquium on 
Quantum Groups and Integrable Systems (Prague, June 18-20, 1998).}}
\vspace*{2cm}
\begin{center}
{\bf Andrzej Frydryszak}\\ 
{\em Institute for Theoretical Physics,
University of Wroclaw, pl. Maksa Borna 9, 50-204 Wroc{\l}aw, Poland}\\ and \\
{\em Laboratoire de Physique Th\'{e}orique, Universit\'{e} Bordeaux I, 33-175 Gradignan,
France}\\
\vspace*{5mm}
{\bf Jerzy Lukierski}\\
{\em Institute for Theoretical Physics,
University of Wroclaw, pl. Maksa Borna 9, 50-204 Wroc{\l}aw, Poland}\\ and \\
{\em Laboratoire de Physique Th\'{e}orique, Universit\'{e} Bordeaux I, 33-175 Gradignan,
France}\\ 
\vspace*{5mm}
{\bf Pierre Minnaert}\\ 
{\em Laboratoire de Physique Th\'{e}orique, Universit\'{e} Bordeaux I, 33-175 Gradignan,
France}\\ 
\vspace*{5mm}
{\bf Marek Mozrzymas}\\
{\em Institute for Theoretical Physics,
University of Wroclaw, pl. Maksa Borna 9, 50-204 Wroc{\l}aw, Poland}\\ and \\
{\em Laboratoire de Physique Th\'{e}orique, Universit\'{e} Bordeaux I, 33-175 Gradignan,
France}
\end{center}
\vspace*{1cm}
\def\bea{\begin{eqnarray}}
\def\eea{\end{eqnarray}}
\def\bel{\begin{equation}\label}
\def\be{\begin{equation}}
\def\ee{\end{equation}}
\def\tens{\otimes}
\def\ba{\begin{array}}
\def\ea{\end{array}}

\begin{abstract}
Three classes of classical $r$ matrices for $sl(4,C)$ algebra are
constructed in quasi-Frobenius algebra approach. They satisfy 
CYBE and are spanned respectively on 8,10,12 generators. 
The $o(4,2)$ reality condition can be imposed only on the eight 
dimensional $r$ matrices with dimension-full deformation parameters.
Contrary to the Poincar\'e algebra case, it appears that all 
deformations with a mass-like deformation parameter ($\kappa$-
deformations) are described by classical $r$-matrices satisfying CYBE.
\end{abstract}
\section{Introduction.}
In four dimensions the deformations of $o(4,2)$ algebra describing
$d=4$ conformal algebra can be obtained by considering the deformations 
of complexified $d=4$ conformal algebra $sl(4,C)$ and
then by taking into account the restrictions imposed by the reality conditions
(which define the deformed $o(4,2)$ algebra as a real form of deformed 
complex $sl(4,C)$).
It appears that the reality conditions are quite restrictive. 
In \cite{lns} there were 
classified all the real forms of Drinfeld-Jimbo deformation $U_q(sl(4,C))$ 
of complexified $d=4$ conformal algebra. It appears that for
standard $*$-Hopf algebra, with $*$-operation being anti-automorphism of 
algebra and  an automorphism of coalgebra
there exist only two real forms of the Drinfeld-Jimbo deformations of 
$U_q(sl(4,C)$, one for $q$ real and second for $|q|=1$, providing the 
$q$-deformations $U_q(o(4,2))$ of $d=4$ conformal algebra.

In this paper we describe deformations of $sl(4,C)$ 
which admit the structure of the real deformed $d=4$ conformal algebra.
To give the classification of quantum deformations,
we shall discuss here the classical $sl(4)$ $r$-matrices. 
We present results which are an essential extension of 
those obtained in \cite{lmm}.
Firstly in sect. 2 we consider the mathematical (Cartan-Weyl for $sl(4)$)
as well as physical (complex-conformal $o(4,2;C)$) basis and then
by providing two possible real 
forms from \cite{lns} we introduce respective physical real 
bases of $d=4$ conformal algebra. 
In sect.\ 3 using the result of \cite{ap} 
and the techniques presented by Alexeevsky, 
Perelomov \cite{ap} and Stolin \cite{sto, sto1} we describe three classes of classical 
$r$-matrices for $sl(4,C)$ spanned on  $8,\, 10,\, 12$ dimensional
subalgebras . 
Considering the $o(4,2)$ real forms we show 
that only one class (eight dimensional) permits the $o(4,2)$ reality 
conditions. It appears that 
these new classical $r$-matrices are span by the generators of $d=4$ Weyl 
algebra (Poincar\'e generators and dilatations)
possibly transformed by Weyl reflections. In Sect. 4 we shall
present some remarks and conclusions (in particular, concerning
the structure relations of the $\kappa$-deformed $d=4$ conformal algebra
by applying the twist transformation proposed by Kulish,
Lyakhovsky and Mudrov \cite{klm}). 
%
\section{Cartan-Weyl basis and its real forms.}
One can write the complex $sl(4;C)$ algebra in Cartan Weyl basis $e_{AB}$
($A,B=1,\ldots,4$); where the choice of indices ($A,B$) is taken from the 
position of non-vanishing entry in the $4\times 4$ fundamental matrix
representation. In particular the diagonal elements $h_1=e_{11}-e_{22}$,
$h_2=e_{22}-e_{33}$, $h_3=e_{33}-e_{44}$ describe three commuting Cartan 
generators, and the simple root generators $e_1=e_{12}$, $e_2=e_{23}$
and $e_3=e_{34}$ describe Cartan-Chevalley basis. 
The composite roots extending 
Cartan-Chevalley basis to Cartan-Weyl basis are described by the 
formulae:
$
e_4\equiv e_{13}=[e_{12}, e_{23}], e_5\equiv e_{24}=[e_{23},e_{34}],
e_6\equiv e_{14}=[e_{12}, e_{24}]=[e_{13},e_{34}].
$
Classical $sl(4)$ algebra is generated by the relations satisfied by the 
Cartan-Weyl basis generators $e_{\pm A}, h_A$ ($A=1,2,\ldots 6$; 
$h_4=h_1+h_2$, $h_5=h_2+h_3$, $h_6=h_1+h_2+h_3$):
\bel{2.2}
\ba{rcl}
[h_A,e_{\pm B}] &=& \pm \alpha _{AB} e_{\pm B}\,,\\
{}[e_i,e_{-j}] &=& \delta_{ij} h_j \, ,\quad i=1,2,3\,,\\ 
{}[e_a,e_{-a}]&=&h_a \, ,\quad  a=4,5,6,
\ea
\ee
Remaining relations of Cartan-Weyl basis of $U(sl(4,C))$ are generated by 
Serre relations and the above definitions. 
In order to describe the real forms of complex Lie algebra $sl(4)$ we 
consider involutive anti-automorphisms $x\to x^*$ of $U(sl(4,C))$ such that for 
any $x,y\in U(sl(4;C))$
\bel{2.7}
(xy)^*=y^*x^* ,\qquad (\mu X +\lambda y)^* = \mu^*x^* +{\lambda}^*y^* ,
\qquad \mu,\, \lambda \in C
\ee
There are the following inequivalent real forms describing by means of the 
reality condition $x=x^*$ the real $o(4,2)$ algebra \cite{lns} ($j=1,2,3$)
\be
h_j^* = -h_{4-j} \, , \qquad  e_{\pm j}^* = e_{\pm (4-j)}
\ee
\be
h_j^* = h_j \, ,\qquad \quad e_{\pm j}^* = \epsilon_j e_{\mp j}  
\ee
with three nonequivalent choices of $(\epsilon_1, \epsilon_2, \epsilon_3)$:
($1,-1,1$), ($-1,1,-1$) and ($-1,-1,-1$).

Let us observe that $sl(4,C) \simeq o(4,2,C)$ and the generators $M_{RQ} = - 
M_{QR}$ ($P,Q=0,1,2,3,4,5$) of $o(4,2;C)$ ($\eta_{AB}=diag(-1,1,1,1,1,-1)$)
satisfy the relations
\bel{2.9}
[M_{PQ},M_{RS}] = \eta_{PS} M_{QR}-\eta_{PR} M_{QS} +\eta_{QR}M_{PS}
-\eta_{QS}M_{PR}
\ee
We extend the Lorentz generators $M_{\mu\nu}=(M_i=\frac12 
\epsilon_{ijk}M_{jk}, L_i=M_{i0})$; $\mu,\nu=0,1,2,3,$ to $d=4$ conformal 
algebra
generators as follows:
\bel{2.10}
P_\mu=(M_{4\mu}+M_{5\mu}) \qquad K_{\mu}=(M_{5\mu}-M_{4\mu})\qquad
D=M_{45}
\ee
The reality conditions lead to the following two ways of 
defining real $d=4$ conformal algebra generators in terms of Cartan-Weyl 
basis of $\hat g=sl(4,C)$; $\hat g=B^-\oplus H\oplus B^-$, where ($B^+,H$), 
($B^-,H$) are two Borel subalgebra and $H=(h_1,h_2,h_3)$ describe Cartan 
subalgebra.\\
i) The reality condition $(B^\pm)^* \subset B^\pm$. 
\bel{2.11}
\ba{rclcrclcrcl}
M_+&=&e_1+e_{-3}&,&M_-&=&-(e_3+e_{-1})&,&M_3&=&\frac{i}{2}(h_1-h3\\
L_+&=&i(e_{-3}-e_1)&,&L_- &=&-i(e_3 - e_{-1})&,&L_3&=&\frac12(h_1+h3)\\
P_1&=&-(e_4+e_5)&,&P_2&=& i(e_4-e_5)&,&P_3&=&i(e_2-e_6)\\
K_1&=&e_{-4} - e_{-5} &,&K_2 &=& i(e_{-4} + e_{-5})&,&K_3&=&i(e_{-2}-e_{-6})\\
P_0&=&-i(e_2+e_6)&,&K_0 &=&i(e_{-2}+e_{-6})&,&D&=&\frac12(h_1+2h_2+h_3)\\
\ea
\ee
where $M_\pm=M_1\pm i M_2,\, L_\pm=L_1 \pm i L_2$.
We see that the Cartan subalgebra $H$ is described by the non compact algebra 
($M_3,L_3,D$), and under the $*$-operation the operators  are
real.\\
ii) The reality condition $(B^{\pm})^*\subset B^\mp$.
This reality condition can
not be applied to the solutions of the CYBE for $sl(4)$ with a dimension of
solution no less then eight.
Hence the assignment of the conformal generators for this case will be not 
needed in further considerations and it is omitted here.\\
It appears that the Cartan subalgebra 
$H$ is described by the compact Abelian subgroup 
($M_{12}=M$, $M_{34}=\frac12(P_3-K_3)$, $M_{50}=\frac12(P_0+K_0)$).
The choices of the generators can be modified if we 
take into consideration the discrete group of Weyl reflections, which 
preserve the Lie-algebra relations. There are three basic Weyl reflections 
$\sigma_1$, $\sigma_2$, $\sigma_3$ describing the automorphism of $sl(4,C)$ 
Lie algebra. For example explicit relations defining $\sigma_1$ are the following:
\bel{2.13a}
\ba{rclcr}
\sigma_1(e_{\pm1})=(a_1)^{\mp1} e_{\mp1}&,&\sigma_1(e_{\pm2}) 
=(a_4)^{\pm1} e_{\pm4}&,&\sigma_1(e_{\pm3})=(a_3)^{\pm1} e_{\pm3} \\
\sigma_1(e_{\pm4})=(a_2)^{\pm1} e_{\pm2}&,&\sigma_1(e_{\pm5}) 
=(a_6)^{\pm1} e_{\pm6} &,&\sigma_1(e_{\pm6})=(a_5)^{\pm1} e_{\pm5}\\
\ea
\ee
$a_4=a_1a_2\quad a_5 = a_2a_3 \quad a_6=a_1 a_2 a_3$.
There exists also the isomorphism of Dynkin diagram 
($\alpha_1\leftrightarrow \alpha _3$) which implies the following 
isomorphism of $sl(4;C)$ Lie algebra:
\bel{2.13d}
\ba{rclclcl}
\beta(e_{\pm1})=e_{\pm3}&,&\beta (e_{\pm2})=e_{\pm2}&,&
\beta(e_{\pm4})=e_{\pm5}&,&\beta(e_{\pm6})=e_{\pm6}
 \ea
\ee
Any product of Weyl reflections 
is again an isomorphism of $sl(4;C)$, but not all these 
isomorphisms commute with the 
$*$-operations defining real forms. The condition
\bel{2.14}
\sigma_{i_1\ldots i_k}\cdot *= * \cdot \sigma_{i_1\ldots i_k}
\ee
is necessary for defining the restriction of Weyl reflections to $o(4,2)$ 
algebra. We obtain
\begin{enumerate}
\item[i)] for the $*$-operation $(B^\pm)^* \subset B^\pm$ the involutions 
$\sigma_2\,,\quad \sigma_1\sigma_3=\sigma_3\sigma_1\,,\quad \beta$
are also isomorphisms of real algebra $o(4,2)$ provided that $b_1^*=b_3$, 
$b_2^*=b_2$.
\item[ii)] for the $*$-operation $(B^\pm)^* \subset B^\mp$ we obtain the following 
isomorphisms of $o(4,2)$: $\sigma_2\,,\beta$. Provided that $b_i^*b_i=1$.
\end{enumerate}
%
\section{Classical $r$-matrices for $sl(4)$ and $o(4,2)$ reality conditions}
We shall consider the antisymmetric solutions of the CYBE i.e.
\be
<<r,r>> = [r_{12},r_{13}]+[r_{12},r_{23}]+[r_{13},r_{23}]=0 \,, ,\qquad 
r\in \hat g \wedge \hat g \,, ,
\ee
where $<<r,r>>$ denotes Schouten bracket ($<<r,r>>\in \hat g \tens \hat g
\tens \hat g$. In order to construct new solutions we shall apply the
quasi-Frobenius algebra approach, using the following definitions and
results \cite{ela, dri, sto, sto1}:
\begin{itemize}
\item[$\diamondsuit$]    Lie algebra $\hat g$ is \underline{quasi-Frobenius}
if there exists a skew-symmetric bilinear form 
$B:\,\,\hat g \wedge \hat g \to C$ such that for all $x,y,x \in \hat g$
\be
B([x,y],z)+B([y,z],x) + B([z,x],y) =0\,.
\ee
Let $b_{ij}=<e_i^* \tens e_j^*, B >$ then $r=r^{ij} e_i \wedge e_j$\, , where
$r^{ij} b_{jk} =\delta_k^i$ satisfies CYBE 
\item[$\diamondsuit \diamondsuit$] If the bilinear form $B$ is determined by a functional
$g_B^*$ on
$\hat g$ such that 
\be
B(x,y) = < g_B^*,[x,y]>\,,
\ee
then $\hat g $ is called \underline{Frobenius} algebra.
\item[$\diamondsuit \diamondsuit \diamondsuit$] The classification of classical
$r$-matrices (obtained in this way) can be reduced to the classification of
quasi - Frobenius algebras which in turn,  
are even dimensional and can be identified with a set of parabolic 
subalgebras.
\end{itemize}
For $\hat g = sl(4,C)$ we can distinguish three relevant families of the
parabolic subalgebras spanning the respective classical $r$-matrices.
Let $B_+=(h_i,\,e_A\,);\,i=1,2,3;\,A=1,\,...\,6$ denotes a Borel subalgebra,
then we have explicitly the following classification of classical
$r$-matrices:\\
{\bf d=12.} Parabolic subalgebra $P_{(-2,-3)}=(B_+ , e_{-2},
e_{-3})$. In this case one obtains the one-parameter generalization of the 
solution given by Gerstenhaber and Giaquinto \cite{gg}
\bea
r_{(12)} = \frac14(3h_1+2h_2+h_3) \wedge e_1 +  \frac14(h_1+2h_2+3h_3)
\wedge e_3
+e_4 \wedge e_{-2}+ 
\nonumber \\
- e_6 \wedge e_{-5} + \lambda ( \frac12(h_1+2h_2+h_3 ) \wedge e_2 + (e_4 +
e_5) \wedge e_{-3})
\eea
This solution of the CYBE has the following properties:
parameter $\lambda$ is arbitrary (it has inverse of mass dimension),
each part of $r$ satisfies CYBE separately, it
does not permit the restriction of $sl(4,C)$ to real $o(4,2)$\\
{\bf d=10.} Parabolic subalgebras $P_{(j)}=(B_+ , e_{-j})$, $j=1,2,3$.
We have to consider here three separate sets of nonsingular functionals:\\
i) Parabolic subalgebra $P_1$.
\be
\ba{rclcrcl}
g_{1a}^*&=&e_5^*+e_4^*+e_1^*&\qquad &g_{1d}^*&=&e_6^*+e_2^*+e_{-1}^*\\
g_{1b}^*&=&e_5^*+e_4^*+e_3^*&\qquad &g_{1e}^*&=&e_6^*+e_2^*+e_1^*+e_3^*\\
g_{1c}^*&=&e_6^*+e_2^*+e_3^*&&&&
\ea
\ee
They yield the following r matrices:\\
 \def\e#1#2{e_{#1} \wedge e_{#2}}
\be
\ba{rcl}
	 r_{1a}^{(10)} &=& -\e23 +\e6{-1} + \frac12 (e_1+e_3)
	 \wedge(h_1+h_3)\\
	  && \frac14 e_4 \wedge (h_1 + 2h_2 -h_3) + \frac14 e_5 \wedge
	  (h_1+2h_2+3h_3) \\
	 r_{1b}^{(10)} &=& -\e12 +\e6{-1} + \frac12 (e_1+e_3)
	 \wedge(h_1+h_3)\\
	  && \frac14 e_4 \wedge (3h_1 + 2h_2 +h_3) + \frac14 e_5 \wedge
	  (-h_1+2h_2+3h_3) \\
	 r_{1c}^{(10)} &=& -\e15 +\e4{-1} + \frac12 (e_3+e_{-1}) 
	                  \wedge(-h_1+h_3)\\
	  && \frac14 e_2 \wedge (-h_1 + 2h_2 +h_3) + \frac14 e_6 \wedge
	  (3h_1+2h_2+h_3) \\
	 r_{1d}^{(10)} &=& -\e15 +\e34 + \frac12 (e_3+e_{-1}) 
	                  \wedge(-h_1+h_3)\\
	  && \frac14 e_2 \wedge (h_1 + 2h_2 -h_3) + \frac14 e_6 \wedge
	  (h_1+2h_2+3h_3) \\
	 r_{1e}^{(10)} &=& -\frac12\e12 +\e6{-1} + \frac14 (e_1+e_3) \\
	                &&  \wedge(h_1+h_3)-\frac12 e_2\wedge(e_3+e_4-e_5)\\
	  && +\frac12 e_4 \wedge (h_1 + h_2) + \frac12 e_5 \wedge
	  (h_2+h_3) \\
\ea
\ee
ii) Parabolic subalgebra $P_{(2)}$.We have shown explicitly by considering the most general ansatz that 
 there does not exist a Frobenius algebra structure on $P_{(2)}^*$\\
iii) Parabolic subalgebra $P_{(3)}$.Nonsingular functionals:\\
\be
\ba{rclcrcl}
g_{3a}&=&e_5^*+e_4^*+e_1^*&\qquad &g_{3d}&=&e_6^*+e_2^*+e_{-3}^*\\
g_{3b}&=&e_5^*+e_4^*+e_3^*&\qquad &g_{3e}&=&e_5^*+e_4^*+e_1^*+e_3^*\\
g_{3c}&=&e_6^*+e_2^*+e_{+1}^*&&&&
\ea
\ee
yielding the following $r$ matrices:\\
\be 
\ba{rcl}              
r_{3a}^{(10)}&=&-\e23 -\e6{-3} +\frac12 (e_1
+e_3) \wedge (h_1 + h_3) \\ \nonumber
&&+ \frac14 e_4 \wedge (h_1+2h_2-h_3) 
+ \frac14 e_5 \wedge (h_1+2h_2+3h_3) \\
r_{3b}^{(10)}&=&-\e12 -\e6{-3} 
+ \frac12 (e_1 +e_3) \wedge (h_1 + h_3) \\
&&+ \frac14 e_4 \wedge (3h_1+2h_2+h_3) 
+ \frac14 e_5 \wedge (-h_1+2h_2+h_3)
\\               
r_{3c}^{(10)}&=&\e34 -\e5{-3} 
- \frac12 (e_1 +e_3) \wedge (-h_1 + h_3) \\
&&+ \frac14 e_2 \wedge (h_1+2h_2-h_3) 
+ \frac14 e_6 \wedge (h_1+2h_2+3h_3)
\\
r_{3d}^{(10)}&=&-\e15 +\e34 
- \frac12 (e_1 +e_{-3}) \wedge (-h_1 + h_3)\\ 
&&+ \frac14 e_2 \wedge (-h_1+2h_2+h_3) 
+ \frac14 e_6 \wedge (3h_1+2h_2+h_3)
\\           
r_{3e}^{(10)}&=&-\frac12 \e15 -\e6{-3}  
+\frac14 (e_1 +e_{3}) \wedge (-h_1 + h_3) \\
&&-\frac12 e_2 \wedge (e_3+e_4-e_5)
+\frac12 e_4 \wedge (h_1 +h_2) 
+ \frac12 e_5 \wedge (h_2+h_3)
\ea
\ee  
It can be shown that all such generated 10-dimensional classical $r$-matrices
do not permit the $o(4,2)$ reality conditions. It is that because
$\sigma_2$ commutes with the $*$ and 
$(\sigma_2 \tens \sigma_2) r_3^{(10)}=r_1^{(10)}$,
but these $r$-matrices are not compatible i.e. 
$<< r_1^{(10)},r_3^{(10)} >> \neq 0$.\\
{\bf d=8.} Here we have the Borel subalgebra $B_+$.
\be
r_1^{(8)} = e_4 \wedge e_3 - e_5 \wedge e_1 + a h_2 \wedge e_6 + 
h_6 \wedge e_6
\ee
Taking into account that above classical $r$-matrix is real 
under the $\ast$-operation (3) and using the Weyl automorphism commuting with
this $\ast$-involution we obtain another form of the d=8 solution: 
\def\s{\sigma}
\bea
r_2^{(8)}=(\sigma_2 \tens \sigma_2) \circ r_1^{(8)}
=[(\s_1\s_3) \tens (\s_1 \s_3) ] \circ r_1^{(8)} = \nonumber \\ 
e_5 \wedge e_{-3} - e_4 \wedge e_{-1} + h_2 \wedge e_2 + a h_6 \wedge e_2
\eea
Let us note that in the physical basis above $r$-matrices are spanned 
on generators of the $d=4$ Weyl subalgebra $(M_i ,L_i ,P_{\mu}, D)$.
\section{Final remarks.}
In this paper we have considered $r$-matrices satisfying CYBE. From their
scaling properties in the physical basis and the fact that invariant three form
for $o(4,2)$; $I=e_{ij}\wedge e_{jk}\wedge e_{ki}\sim M_A^B \wedge M_B^C
\wedge M_C^A$ is scale invariant it follows that every $r$-matrix giving 
dimension-full
deformation ($\kappa$-deformation) satisfies CYBE.\\ 
We do not present here the description of the complete $\kappa$-deformed
algebra. Using the results of the work \cite{klm} one can obtain the 
coproduct applying to the ${\Delta}_0$\ the twist $F$ of the form: 
$
{\Delta}_F(x) = F\circ {\Delta}_0(x) \circ F^{-1}
$
, where
$
F=\exp{(h_6 \tens \s (e_6 ))}\cdot \exp{(2\lambda  e_1 \tens e_5 
\cdot e^{-2\s (e_6 )})}\cdot
\exp{(2\lambda e_4 \tens e_3 e^{-2\s (e_6 )})}
$
and
$
\s (e_6 )= -\frac12 
\ln{(1-2\lambda e_6 )} \sim \ln{(1-\frac{i}{\kappa}(P_0 +P_3 ))}
$
. It will be given in a forthcoming paper of the present authors.\\

\end{document}